\documentclass{article}
\usepackage{graphicx}  
\usepackage{amsmath}   
\usepackage[compress]{cite}
\usepackage{amssymb}   
\usepackage{bm} 
\usepackage{dcolumn}
\usepackage{color}
\usepackage{mathrsfs}
\usepackage{amsfonts}
\usepackage{varioref}
\usepackage{textcomp}
\usepackage[titletoc]{appendix}
\RequirePackage[colorlinks,citecolor=blue,urlcolor=magenta,linkcolor=blue]{hyperref}
\allowdisplaybreaks
\def\LL{Lanczos-Lovelock }

\newcommand{\nn}{\nonumber}
\labelformat{section}{Section #1} 
\labelformat{subsection}{Section #1} 
\labelformat{subsubsection}{Section #1}
\labelformat{subsubsubsection}{Section #1}
\labelformat{equation}{Eq.~(#1)} 
\labelformat{figure}{Fig.~#1} 
\labelformat{subfigure}{Fig.~\thefigure#1} 
\labelformat{table}{Tab.~#1} 
\labelformat{appendix}{Appendix #1}
\title{Eddington gravity with matter: An emergent perspective}
\author{Sumanta Chakraborty\footnote{sumantac.physics@gmail.com}$~^{1}$ 
and T. Padmanabhan\footnote{paddy@iucaa.in}$~^{2}$\\
{$~^{1}$\small{School of Physical Sciences}}\\
{\small{Indian Association for the Cultivation of Science, Kolkata-700032, India}}\\
{$~^{2}$\small{IUCAA, Post Bag 4, Ganeshkhind, Pune 411007, India}}}
\begin{document}
  
\maketitle
\begin{abstract}
We describe an action  principle, within the framework of the Eddington gravity, which incorporates the matter fields in a simple manner. Interestingly, the gravitational field equations derived from this action is identical to the Einstein's equations, in contrast with the earlier attempts in the literature. The cosmological constant arises as an integration constant in this approach. In fact, the derivation of the field equations \textit{demands} the existence of a non-zero cosmological constant, thereby providing the \textit{raison d'\^{e}tre} for a non-zero cosmological constant, implied by the current observations.  Several features of our approach strongly support the paradigm that gravity is an emergent phenomenon and, in this perspective, our action principle could have a possible origin in the microstructure of the spacetime. We also discuss several extensions of the action principle, including the one which can incorporate torsion in the spacetime. We also show that an Eddington-like action can be constructed to obtain the field equations of the Lanczos-Lovelock gravity.
\end{abstract}

\section{Introduction and Motivation}

The usual starting point for obtaining the  gravitational field equations in general relativity is the Hilbert action, with the metric treated as the dynamical variable. Even though it is possible to derive the Einstein's equations from the variation of this  action (with respect to the metric), it is not completely straightforward because  the variational problem  is ill-posed. This is because of the presence of second derivatives of the metric in the Ricci scalar, which is the Lagrangian for the Hilbert action \cite{Padmanabhan:2014lwa,Charap:1982kn,Gibbons:1976ue,York:1972sj,Dyer:2008hb}. The well-posed version of the Hilbert action  can only be obtained by adding suitable boundary terms to the Hilbert action, which crucially depend on the choice of the boundary surface \cite{Parattu:2015gga,Chakraborty:2016yna,Parattu:2016trq,Jubb:2016qzt,Lehner:2016vdi,Chakraborty:2018dvi,Chakraborty:2017zep}. Though these boundary terms do not affect the Einstein's equations, they have important thermodynamical as well as geometrical implications \cite{Chakraborty:2019doh,Padmanabhan:2019art,Padmanabhan:2013nxa}. 

Such complications, associated with the Hilbert action, provides one possible motivation for an alternative proposal, originally suggested by Eddington \cite{Eddington:1924,Schrod:1950}. In this approach, one takes the gravitational Lagrangian to be $\sqrt{\textrm{det.}(R_{ab})}$, where $R_{ab}$ is the Ricci tensor. Since the Ricci tensor $R_{ab}$ (in the (0,2) form) can be constructed solely from the connection $\Gamma^{p}_{qr}$, we can consider the \textit{connection} to be the dynamical variable and vary it in the Eddington action. Interestingly, the outcome is the Einstein's equations with a \emph{non-zero} cosmological constant. Thus Eddington's action is a viable alternative to the Hilbert action and, as we will describe below, is also well-posed.  

There is, however, a major issue with the Eddington's proposal, viz. that it does not include the matter degrees of freedom. Surprisingly, the inclusion of matter in the Eddington's framework has turned out to be not very straightforward. Most of the proposals in the literature are motivated by the Born-Infeld like structure \cite{Born:1934gh,Poplawski:2006zr,Delsate:2012ky,Vollick:2003qp,Avelino:2012ge,Deser:1998rj,Banados:2008fj,Pani:2011mg,Banados:2010ix}. Broadly speaking, this requires the action to be dependent on the connection $\Gamma^{p}_{qr}$, the matter degrees of freedom $\Psi$ and the metric $g_{ab}$, though the connection and the metric are considered independent. Setting the variation of the action with respect to the metric to zero, i.e., $(\delta \mathcal{A}[\Gamma,g,\Psi]/\delta g=0)$, one obtains $g=g(\Gamma,\Psi)$. This result, when substituted back to the action yields the on-shell action, $\mathcal{A}[\Gamma,\Psi]$, which is a function of the connection and the matter fields alone. The final step is the variation of this action with respect to the connection which yields the desired gravitational field equations. However, none of these attempts, as far as we know, yields the Einstein's equations; rather they lead to  additional corrections \cite{Vollick:2003qp,Vollick:2005gc,Vollick:2006qd}.

In this paper, we will discuss a completely new approach to derive the Einstein's equations in the spirit of Eddington gravity. Most importantly,  we will construct an action principle such that its variation, with respect to the connection, leads precisely to the Einstein's equations \emph{without} any additional corrections. In addition, as we will demonstrate, our variational principle  will be well-posed.\footnote{Another key motivation for this approach is the following. The action we will be using here is closely related to an effective action in the emergent gravity paradigm,  when we integrate out certain microscopic degrees of freedom of spacetime in a path integral. We will not pursue this idea in this paper --- except for a brief comment in the last section and in \ref{AppE2} --- but will discuss it in a separate work \cite{TPSC2020}.} 

The paper is organized as follows: In \ref{Mat_Ed_Action}, the modified action for the gravity (incorporating the matter field) along the lines  of Eddington gravity is presented. Its variation, leading to the gravitational  field equation, is described in \ref{Mat_Ed}. Finally the relationship of this action with the microstructure of spacetime is briefly discussed in \ref{conclusion}. The Appendices contain the mathematical details and possible further extensions.

\textit{Notations and Conventions:} We will assume $c=1=\hbar$ and use the mostly positive signature convention. We will work in $d$ dimensional spacetime except when specified otherwise. Latin sub/superscripts run over all the spacetime indices.
\section{Action for Eddington gravity with matter}\label{Mat_Ed_Action}

In this section, we will introduce our  action principle for  gravity plus  matter and describe some of its key features.  We will vary this action and derive the field equations in the next section. 

Motivated by the original form of the Lagrangian associated with the Eddington gravity, we propose the following action, in $d>2$ spacetime dimensions, to describe gravity coupled to matter:
\begin{align} \label{action}
\mathcal{A}=\int d^{d}x~\sqrt{|\textrm{det.}\left(R_{(ab)}(\Gamma)-\kappa \bar{T}_{ab}\right)|}~;\qquad \bar{T}_{ab}=T_{ab}-\frac{1}{(d-2)}Tg_{ab}~,
\end{align}
where, $\kappa=8\pi G$, with $G$ being the Newton's gravitational constant. There are several features of this action which are worth emphasizing.

(i) The action given above (with an integration measure $d^{d}x$ rather than $\sqrt{|\textrm{det.} g|}~d^{d}x$) is indeed a scalar, since the determinant of any second rank tensor field transforms identically to $\sqrt{|\textrm{det.} g|}$, thereby making the action a generally covariant scalar. The action is also dimensionless. These features, of course, should be obvious to those familiar with the standard Eddington gravity, which is obtainable from \ref{action} by setting $T_{ab}=0$.

(ii) In the above action, $R_{(ab)}(\Gamma)\equiv (1/2)[R_{ab}(\Gamma)+R_{ba}(\Gamma)]$ is the symmetric part of the Ricci tensor, which is constructed solely from the connection $\Gamma^{a}_{bc}$. Thus the gravitational sector is independent of the metric and depends only on the connection. For simplicity, in the main discussion, we will assume that the connection is symmetric , i.e.,  $\Gamma^{a}_{bc}=\Gamma^{a}_{cb}$. This assumption can be relaxed --- and torsion can be included through the antisymmetric part of the connection--- rather easily, as demonstrated in \ref{AppC}. 

(iii) Since we are not assuming any a priori relation  between $\Gamma^{a}_{bc}$ with the metric $g_{cd}$, it follows that the Ricci tensor need not be symmetric. This is due to the term $\partial_{a}\Gamma^{c}_{bc}$ in the Ricci tensor, which is not symmetric in $(a,b)$ unless $\Gamma^{a}_{bc}$ is given by the Christoffel symbol corresponding to a metric. Therefore, we have constructed the action out of the symmetric part of the Ricci tensor.

(iv) The  $T_{ab}$ is the matter energy momentum tensor, whose trace is denoted by $T$. This can --- and indeed it does --- depend on the metric. However, we will assume that  $T_{ab}$ is independent of the connection; this criterion is satisfied by almost all the matter stress tensors we will be interested in. As already stated, the connection and the metric are treated as independent variables at this stage, somewhat in the spirit of the Palatini formulation in standard Einstein's theory. (The assumption that matter sector does not explicitly depend on the connection is an assumption usually made in the standard Palatini approach as well.)

(v) The most significant departure of the action principle proposed above, from those in the previous literature is in the treatment of the matter degrees of freedom. Instead of working with a separate matter Lagrangian, we will work with the matter energy-momentum tensor $T_{ab}$ itself. Though it may appear somewhat surprising at first sight, everything will work out satisfactorily due to the following two facts: First,  given a Lagrangian for the matter field, the energy momentum tensor can be uniquely determined and hence there is a clear correspondence between the two. Second, as we shall show, the gravitational field equations will lead to $\nabla_a T^a_b=0$ (as in the standard Einstein's theory) from which one can derive the equations of motion for the matter field. To reiterate, variation of the above action with respect to the connection $\Gamma^{a}_{bc}$ will yield the gravitational field equations, which, as we will demonstrate in the next section, will be identical to the Einstein's equations sourced by $T_{ab}$ (and a cosmological constant). Then Bianchi identity will yield the field equations for the matter field through $\nabla_a T^a_b=0$. (We will comment again on this aspect later, after the derivation of the field equations have been presented.)  

We conclude this section with some brief comments on the relation between a second-rank tensor and a matrix.  Given any second rank tensor ${S}_{ab}$ (equivalently ${S}^{a}_{b}$, or, ${S}^{ab}$), one can construct a matrix $\mathcal{M}^{a}_{~b}$, such that the $a$th row and the $b$th column of the matrix coincides with the $(a,b)$th element of the tensor (given in (0,2), (1,1) or (2,0) form). The determinant of the tensor is then \textit{defined} as the determinant of the matrix to which its components are mapped.\footnote{Some pedagogical subtleties in defining the determinant of an arbitrary second rank tensor, not adequately emphasized in the literature and textbooks, are discussed in \ref{AppA}.} This is precisely what we do while computing the determinant of the metric tensor $g_{ab}$ in general relativity;  the determinant in \ref{action} is computed exactly as we compute the determinant of the metric tensor $g_{ab}$ in general relativity. (Once again, this should be clear to those who are familiar with standard Eddington gravity, which is obtainable from \ref{action} on setting $T_{ab}=0$.)

\section{Variation of the action and the gravitational field equations}\label{Mat_Ed}

We will now vary the action in \ref{action} for arbitrary variation of the connection and shall obtain the gravitational field equations. For this purpose, it will be convenient to define the  tensor $M_{ab}\equiv R_{(ab)}-\kappa \bar{T}_{ab}$, which, by construction, is symmetric. As mentioned earlier, the definition of the determinant appearing in \ref{action}, requires us to map the (components of) tensor $M_{ab}$ to (the elements of) a matrix $\mathcal{M}^{a}_{~b}$. Given the matrix $\mathcal{M}^{a}_{~b}$, one can define the inverse matrix $\mathcal{N}^{a}_{~b}$, such that, $\mathcal{M}^{a}_{~b}\mathcal{N}^{b}_{~c}=\delta^{a}_{c}=\mathcal{N}^{a}_{~b}\mathcal{M}^{b}_{~c}$. Again, one can map the inverse matrix $\mathcal{N}^{a}_{~b}$ back to a tensor $N^{ab}$, such that $N^{ab}M_{bc}=\delta^{a}_{c}=M_{cb}N^{ba}$. (For some pedagogical details, see \ref{AppA}.) This tensor $N^{ab}$ will be useful in the ensuing analysis. 

The variation of our action in \ref{action}, under arbitrary variation of the symmetric connection $\Gamma^{a}_{bc}$, leads to:
\begin{align}
\delta \mathcal{A}&=\int d^{d}x~\frac{1}{2\sqrt{|\textrm{det.}\left(\mathcal{M}\right)|}}|\textrm{det.}\left(\mathcal{M}\right)|
\times \mathcal{N}^{a}_{~b}\delta \mathcal{M}^{b}_{~a}
\nonumber
\\
&=\frac{1}{2}\int d^{d}x~\sqrt{|\textrm{det.}\left(\mathcal{M}\right)|}~N^{ab}\delta M_{ba}~,
\end{align}
where, $\textrm{det.}(\mathcal{M})$ denotes the determinant of the matrix $\mathcal{M}^{a}_{b}$. Since the matter energy-momentum tensor $\bar{T}_{ab}$ is independent of the connection,  the variation of the tensor $M_{ab}$, due to an arbitrary variation of the connection, will arise only from the  Ricci term. The variation of the Ricci tensor, due to variation of the connection, is given by:
\begin{align}
\delta R_{(ab)}=\nabla_{c}\delta \Gamma^{c}_{ab}-\nabla_{(a}\delta \Gamma^{c}_{b)c}~,
\end{align}
So the variation of the action $\mathcal{A}$ becomes, 
\begin{align}\label{variation_01}
\delta \mathcal{A}&=\frac{1}{2}\int d^{d}x~\sqrt{|\textrm{det.}\left(M\right)|}~N^{ba}\left(\nabla_{c}\delta \Gamma^{c}_{ab}-\nabla_{a}\delta \Gamma^{c}_{bc}\right)
\nonumber
\\
&=\frac{1}{2}\int d^{d}x~\sqrt{|\textrm{det.}\left(M\right)|}~\delta_{ab}^{cd}~N^{pa}\nabla_{d}\delta \Gamma^{b}_{cp}~;\qquad \delta^{ab}_{cd}\equiv \delta^{a}_{c}\delta^{b}_{d}-\delta^{a}_{d}\delta^{b}_{c}~,
\end{align}
where we have used the fact that $N^{ab}$ is symmetric. In deriving the above variation of the action we have adopted the usual convention of writing the determinant of a matrix, $\textrm{det.}(\mathcal{M})$, as the determinant of the tensor, $\textrm{det.}(M)$, since no confusion is likely to arise in the subsequent discussion. 

The above variation can be simplified further. As a first step,  we will rewrite the above expression by separating out a total derivative term:
\begin{align}\label{variation_01N}
\delta \mathcal{A}&=\frac{1}{2}\int d^{d}x~\nabla_{d}\left[\sqrt{|\textrm{det.}\left(M\right)|}~\delta_{ab}^{cd}~N^{pa}\delta \Gamma^{b}_{cp}\right]-\frac{1}{2}\int d^{d}x~\nabla_{d}\left[\sqrt{|\textrm{det.}\left(M\right)|}~\delta_{ab}^{cd}~N^{pa}\right]\delta \Gamma^{b}_{cp}~.
\end{align}
We next want to convert the total divergence term to a surface term, as is usually done, so that it will vanish with the usual boundary conditions, viz. $\delta \Gamma^{b}_{cp}=0$ at the boundary. This is, of course, trivial if the expression had a $\sqrt{|\textrm{det.}\left(g\right)|}$ in place of $\sqrt{|\textrm{det.}\left(M\right)|}$ in the first integrand. Due to the presence of the $\sqrt{|\textrm{det.}\left(M\right)|}$ factor, it may appear that such a conversion of the total divergence term to a boundary term will not be possible in the present context. Fortunately, it turns out that one can indeed convert the first term in \ref{variation_01N} into a surface term because the following identity holds (see \ref{AppB} for a derivation):
\begin{align}\label{identity}
\nabla _{c}\left[\sqrt{|\textrm{det.}\left(M\right)|}~V^{c}\right]=\partial_{c}\left[\sqrt{|\textrm{det.}\left(M\right)|}~V^{c}\right]~,
\end{align}
for any vector field $V^{c}$ and second rank tensor field $M_{ab}$, whose determinant is $\textrm{det.}(M)$. Using this result, the variation of the action  in \ref{variation_01N} becomes,
\begin{align}
\delta \mathcal{A}&=\frac{1}{2}\int d^{d}x~\partial_{d}\left[\sqrt{|\textrm{det.}\left(M\right)|}~\delta_{ab}^{cd}~N^{pa}\delta \Gamma^{b}_{cp}\right]-\frac{1}{2}\int d^{d}x~\nabla_{d}\left[\sqrt{|\textrm{det.}\left(M\right)|}~\delta_{ab}^{cd}~N^{pa}\right]\delta \Gamma^{b}_{cp}~.
\end{align}
The first term provides the boundary contribution arising out of the action when it is varied with respect to the connection, while the second term will provide the gravitational field equations. Even though the boundary term will not contribute to the field equations, it is worth emphasizing a few points about the boundary contribution, as it is intimately connected with the question of whether the action principle is well-posed. Let us take the usual boundary of a four dimensional volume, made out of two constant time hypersurfaces, $t=t_1,t=t_2 $, along with a timelike surface at spatial infinity. Then we will be fixing $\Gamma^{a}_{bc}$ at both the $t=t_1,t=t_2 $ hypersurfaces (and assume that $\delta\Gamma^{a}_{bc}$  vanishes at spatial infinity). Thus, in the present context, the field equations must be of second order in $\Gamma^{a}_{bc}$, for the variational problem to be well-posed. As evident from the second term of the above variation, the field equations depend on at most the second derivatives of the connection and hence the variational problem is indeed well-posed. This is unlike the \textit{metric} variation of the Hilbert action, in which case not only the metric, \textit{but also} its normal derivatives need to be fixed at the boundaries.  Thus the action for Eddington gravity will not require any additional boundary term, in sharp contrast to the Hilbert action. 

Neglecting the boundary contribution and setting $\delta \mathcal{A}=0$ for arbitrary variation of the connection in the bulk, we obtain the field equations to be:
\begin{align}\label{mod_field_eqN}
\nabla_{d}\left[\sqrt{|\textrm{det.}\left(M\right)|}~\delta_{ab}^{cd}~N^{pa}\right]&+\nabla_{d}\left[\sqrt{|\textrm{det.}\left(M\right)|}~\delta_{ab}^{pd}~N^{ca}\right]
\nonumber
\\
&\hskip -3 cm =2\nabla_{b}\left[\sqrt{|\textrm{det.}\left(M\right)|}~N^{pc}\right]-\delta^{c}_{b}\nabla_{d}\left[\sqrt{|\textrm{det.}\left(M\right)|}~N^{pd}\right]-\delta^{p}_{b}\nabla_{d}\left[\sqrt{|\textrm{det.}\left(M\right)|}~N^{cd}\right]=0~.
\end{align}
(In writing the first line, we have taken care of the fact that $\delta \Gamma^{b}_{cp}$ is symmetric in $c,p$.) This equation can also be simplified further, by multiplying both sides using $\delta^{b}_{c}$, from which we immediately obtain, $\nabla_{a}[\sqrt{|\textrm{det.}\left(M\right)|}~N^{ba}]=0$. Substituting this expression back in \ref{mod_field_eqN}, we finally obtain,
\begin{align}\label{mod_field_eqF}
\nabla_{c}\left[\sqrt{|\textrm{det.}\left(M\right)|}N^{ab}\right]=0~.
\end{align}
This provides the gravitational field equations, arising out of the connection $\Gamma^{a}_{bc}$ and is indeed a second order differential equation in $\Gamma^{a}_{bc}$, leading to a well-posed boundary value problem, as mentioned before. 

If we set $T_{ab}=0$, so that $M_{ab}=R_{(ab)}$, this equation reduces to the one we would have obtained in the context of standard Eddington gravity. In the current context, with the presence of matter, this equation has the same structure as that in the Eddington gravity, with $R_{(ab)}$ replaced by $M_{ab}$. So we can proceed exactly as in the case of Eddington gravity, as should be obvious to those familiar with the standard Eddington gravity analysis. Nonetheless, we will spell out the relevant algebraic details below. 

The field equation arising from the variation of the connection, given by \ref{mod_field_eqF}, contains second derivatives of the connection. However, as in standard Eddington gravity, this equation  can be immediately integrated to give the first integral (which will involve only the first derivatives of the connection). To do this we only have to note that \ref{mod_field_eqF} requires us to find a second rank, symmetric tensor density which has vanishing covariant derivative with respect to the connection $\Gamma^{a}_{bc}$, which is used to define the derivative operator $\nabla_c$. Because $N^{ab}$ and $M_{ab}$ are inverses of each other, \ref{mod_field_eqF} requires $\nabla_c M_{ab}=0$. If we expand out the covariant derivative in this equation, we can express $\Gamma^{a}_{bc}$ in terms of $N^{ab}$ and the  derivatives of $M_{ab}$, exactly as we would relate the metric to the connection using the condition $\nabla_c g_{ab}=0$. Therefore, when we set $M_{ab}\propto g_{ab}$  the connection $\Gamma^{a}_{bc}$ and  the metric will be related in the standard manner. (That is, the connection used in $\nabla_c$ will be the one compatible with the metric.) This will also, in turn, make  $N^{ab} \propto g^{ab} $ and $ \textrm{det.}(M) \propto \textrm{det.}(g)$. With this choice \ref{mod_field_eqF} reduces to $\nabla_{c}[\sqrt{|\textrm{det.}(g)|}g^{ab}] =0$, which is just the standard compatibility condition between the metric and  $\Gamma^{a}_{bc}$ used to construct the connection, thereby closing the logical loop. (This is exactly the same as  what is done in the standard Eddington gravity, when $T_{ab}=0$.) 

Before proceeding further, we will mention one subtlety in the above argument in the current context, when $T_{ab}\neq0$. This is related to an interesting and hidden role played by the principle of equivalence in the presence of matter. Suppose we introduce \textit{some} metric $q_{ab}$ from which  the connection $\Gamma^{a}_{bc}$ can be obtained in the standard manner. Then the  first integral to \ref{mod_field_eqF} is indeed given by:
\begin{align}
\sqrt{|\textrm{det.}\left(M\right)|}N^{ab}\propto\sqrt{|\textrm{det.}\left(q\right)|}q^{ab}~,
\label{FF_eq_mod}
\end{align}
where $q^{ab}$ is the inverse of the metric tensor $q_{ab}$. At this stage, \textit{formally}, we actually have  \textit{two} metric tensors in play: $g_{ab}$, which could occur in the matter  sector  of the action (through $\bar{T}_{ab}$) and the tensor $q_{ab}$, which is introduced as the one compatible with the connection $\Gamma^{c}_{ab}$, and arises in the gravitational sector through \ref{FF_eq_mod}. The principle of equivalence, however, requires us to identify these two metric tensors (i.e., set $q_{ab}=g_{ab}$) and make $\Gamma^{a}_{bc}$  the Christoffel symbol associated with either of them. To see this, note that principle of equivalence allows us to choose a coordinate system around any event $\mathcal{P}$ such that the local physics reduces to that of special relativity and all gravitational effects vanish to first order. This, in  turn, is possible only if we can choose a coordinate system such that the metric reduces to the Minkowski form ($\eta_{ab}$) at $\mathcal{P}$ and the Christoffel symbols derived from the metric vanish at  $\mathcal{P}$. Such a choice of coordinate system is clearly not possible if there are \textit{two} non-trivially different metrics $g_{ab}$ and $q_{ab}$ (as well as their corresponding connections). A single coordinates transformation will not be able to reduce two non-trivially different metrics into locally flat form, simultaneously. Since we want \textit{both} the geometrical effects governed by $q_{ab}$ and the behaviour of matter governed by $g_{ab}$ to reduce to special relativistic form in the \textit{same} freely falling frame, it is necessary that we identify $g_{ab} =q_{ab}$. (A more general class of theories, called bimetric theories of gravity,  is possible if we relax this condition. We will not be concerned with such generalizations in this work.)

Returning to the main discussion, the first integral to \ref{FF_eq_mod} leads to the identification $M_{ab}\propto g_{ab} \equiv \lambda g_{ab}$ where $\lambda$ is an integration constant. The introduction of the metric, compatible with the connection, also makes the Ricci tensor symmetric $R_{ab}=R_{ba}$ so that the equation $M_{ab} = \lambda g_{ab}$, leads to:
\begin{align}
R_{ab}-\kappa\bar{T}_{ab}=\lambda g_{ab}~. \label{ricci}
\end{align}
Taking the trace of this equation, we obtain, on the left hand side, $R+(2/(d-2))\kappa T$ and on the right hand side, $\lambda d$, thereby yielding, $R=\lambda d-(2/(d-2))\kappa T$. So, with some simple algebra, \ref{ricci} can be rewritten, in terms of the Einstein tensor $G_{ab}\equiv R_{ab}-(1/2)Rg_{ab}$ as:
\begin{align}
G_{ab}+\left(\frac{d-2}{2}\right)\lambda g_{ab}
\equiv G_{ab}+\Lambda g_{ab}
=\kappa T_{ab}~. \label{Einstein_eq}
\end{align}
where, $\Lambda\equiv [(d/2)-1]\lambda$ is the $d$-dimensional cosmological constant. So we have obtained precisely the Einstein's equations with a cosmological constant term. In contrast to other approaches to include matter in the Eddington gravity, where the gravitational field equations themselves are modified, while in the present approach we obtain the exact Einstein's equations. 

\subsection{Comments on the result}

We will now make several key comments about our result.

(a) The role played by the cosmological constant in the above derivation is note-worthy. In sharp contrast to the standard derivation of the Einstein's equations from the Hilbert action, our approach \textit{demands} the existence of a non-zero cosmological constant. In order to arrive at the above field equation we \textit{must} have $\Lambda \neq 0$, i.e., a non-zero cosmological constant is absolutely necessary if the field equation arising out of the Eddington gravity has to make any sense. This is, of course, gratifying because in the usual approaches, the cosmological constant is an ``optional'' parameter; you can set it to zero value or to a non-zero value, as desired. Within such an approach, the observational facts, indicating the existence of a non-zero cosmological constant, have no fundamental explanation. That is, the standard derivation of the Einstein's equation from the Hilbert action goes through without any hitch even if the cosmological constant is zero. But in our approach, the derivation of the field equations \textit{demands a non-zero cosmological constant}.

(b) The cosmological constant arises in the  first integral to the equations of motion in the form of an integration constant. It has been stressed in the previous literature that this is indeed the only way the cosmological constant problem can be addressed (see e.g., \cite{Padmanabhan:2013nxa}). This is reminiscent of the manner in which cosmological constant arises in emergent gravity paradigm (see e.g., \cite{Padmanabhan:2019art}). Just as in these approaches, its numerical value has to be fixed using some other general principle (see e.g., \cite{Padmanabhan:2017qvh}), since it is an integration constant. We will say more about the connection between this approach and emergent gravity paradigm in the last section.

(c) We never varied the metric in the action principle and only needed to vary the connection. The resulting equations of motion, \ref{mod_field_eqF} involves second derivatives of the connection; but it was trivial to find the first integral to this equation of the form $M_{ab}=\lambda g_{ab}$ which involves only the first derivatives of the connection and hence the second derivatives of the metric. All these are true  even in the standard approach to Eddington gravity. The only  modification is that, in the absence of matter, the action in \ref{action} is independent of the metric, while our action has a possible dependence on the metric though the energy momentum tensor $T_{ab}$. The variational principle treats the connection and the metric as independent variables and we have \textit{only} varied the connection. Of course, just because some quantity appears in the action does \textit{not} mean that  we must vary it. More formally, the complete specification of \textit{any} variational principle has three ingredients: (i) The form of the action functional. (ii) The entities which are varied. (iii) The nature of the variation and the boundary conditions. We have complete liberty to prescribe these; as long as we choose a consistent set of these three ingredients, we do have a valid variational problem, as in our case. 

(d) Closely related to the above comment is the \textit{dual} role played by the field equations, as in \ref{mod_field_eqF}. Its first integral, given by $M_{ab}=\lambda g_{ab}$, achieves \textit{two} things: First of all, this choice transforms \ref{mod_field_eqF} to read $\nabla_c(\sqrt{|\textrm{det.}g|}g^{ab})=0$, which is the standard compatibility condition between the metric and the connection; so we obtain this without having to vary the metric, unlike, say, in the standard Palatini approach. (This happens in the usual Eddington gravity without matter as well.) Second, using $M_{ab}=R_{ab}-\kappa T_{ab}$, we obtain the Einstein's equations, in the form $R_{ab}-\kappa T_{ab}=\lambda g_{ab}$. The original equation, i.e., \ref{mod_field_eqF}, which is second order in the connection would suggest that the connection is the dynamical variable of the theory; but once we obtain the first integral, $R_{ab}-\kappa T_{ab}=\lambda g_{ab}$, which is first order in the connection but second order in the metric, we see that the metric acquires the status similar to dynamical variable, even though we never needed to vary the metric in the action principle. So the metric becomes, in the Wheelerian language a ``dynamical variable without (being) a dynamical variable.'' This is not only consistent with emergent gravity paradigm but even strongly suggests thinking of metric as an emergent variable.

(e) The Bianchi identity immediately gives $\nabla_{a} T^{a}_{b}=0$, which is a consistency condition on $T^{a}_{b}$, which is used in the action. In addition, this will lead to the equations of motion for the matter field without having to vary the matter variables separately. (``Spacetime tells matter how to move''.) This is perfectly adequate in a completely classical theory, in which the action is just a tool to get the equations of motion incorporating the relevant symmetries in the most economical way. However, there could be some contexts, like in the study of quantum fields in a fixed curved geometry, in which we would like to get the equations of motion for the matter sector from variation of matter variables in the total action. The same issue arises in the emergent gravity paradigm as well (see the discussion in the third para after Eq. (40) in \cite{Padmanabhan:2007xy}) and can be taken care of by the following prescription. The total action is  taken to be the sum of $L_{\rm matter}$ and the action in \ref{action}. The connection is varied first and the solution to the gravitational field equations is substituted into the action to obtain the on-shell action as far as gravity is concerned. The matter degrees of freedom are then varied in this on-shell action to obtain matter equations of motion;  one can also perform a path integral over matter variables in the on-shell action to do standard quantum field theory in curved spacetime. Note that the gravitational part of the on-shell action will be the one obtained by replacing,  $\textrm{det.}(R_{(ab)}-\kappa\bar{T}_{ab})$ by $\lambda^{d}\det.(g)$, which is devoid of any matter degrees of freedom. Thus the new on-shell action will involve only the matter Lagrangian $L_{\rm matter}$ plus an additional metric dependent term, proportional to $\lambda^{d/2}$. Therefore, the variation of the matter degrees of freedom, is  identical to the variation of the matter Lagrangian $L_{\rm matter}$, in a given curved spacetime and will lead to the correct evolution equation for the matter fields.

(f) We next comment on the case of non-vanishing torsion within the context of the Eddington gravity. The first step, again, is to choose the appropriate Lagrangian and thus the action. Following the previous discussion, it seems legitimate to consider the Lagrangian to depend on the Ricci tensor constructed out of the symmetric Christoffel connection and the contorsion tensor, i.e., one may again consider the following Lagrangian $\sqrt{|\textrm{det.}(\bar{R}_{(ab)}-8\pi G\bar{T}_{ab})|}$. Here, $\bar{R}_{ab}$ is the Ricci tensor which depends on the spacetime torsion as well and hence is not symmetric. But it turns out that the variation of the above action with respect to the symmetric Christoffel connection and the contorsion tensor does not yield appropriate expressions for the gravitational field equations. However, as  shown in \ref{AppC}, a specific modification of the action does yield the correct gravitational field equations for the Einstein-Cartan theory. Of course, in the absence of any Fermionic or non-minimally coupled matter field, the torsion tensor will vanish identically on-shell and then the Einstein-Cartan gravitational field equations will reduce to the Einstein's equations.

(g) Once we have broken free from the compulsion to vary the metric, it is possible to construct the Eddington-type action for \LL models (coupled to matter) as well. (For a review of \LL models, see \cite{Padmanabhan:2013xyr}). In this case, for the $m$th order \LL gravity,  we take the action to be:
\begin{align} 
\mathcal{A}&=\int d^{d}x~\sqrt{|\textrm{det.}\left(\mathcal{R}^{(m)}_{(ab)}-\kappa \bar{T}^{(m)}_{ab}\right)|}~;\qquad \bar{T}^{(m)}_{ab}=T_{ab}-\frac{1}{(d-2m)}Tg_{ab}~;
\end{align}
where,
\begin{align}
\mathcal{R}^{(m)}_{ab}&=P_{a}^{~pqr}R_{bpqr}=m~\delta^{c_{1}d_{1}c_{2}d_{2}\cdots c_{m}d_{m}}_{a_{1}b_{1}a_{2}b_{2}\cdots a_{m}b_{m}}\left(g^{b_{2}e_{2}}R^{a_{2}}_{~~e_{2}c_{2}d_{2}}\cdots g^{b_{m}e_{m}}R^{a_{m}}_{~~e_{m}c_{m}d_{m}} \right)\delta^{a_{1}}_{a}g^{b_{1}e_{1}}g_{bq}R^{q}_{~e_{1}c_{1}d_{1}}~.
\end{align}
Treating the metric and the connection as independent and varying the connection, somewhat lengthy algebra (see \ref{AppD} for details) lead to the standard field equations for the \LL model:
\begin{equation}
 \mathcal{R}_{(ab)}^{(m)}-\frac{1}{2}L_{(m)}g_{ab}+\Lambda g_{ab}=\kappa T_{ab}; \qquad \Lambda=\left(\frac{d-2m}{2m}\right)\lambda~.
\end{equation}
Note that for $d=2m$, i.e., in the critical dimension for the $m$-th order pure Lovelock gravity, the effect from the cosmological constant term identically vanishes.

\section{Discussion: Extensions and the broader perspective}\label{conclusion}

We have explicitly demonstrated that there exists a first order formalism, in the same spirit as the Eddington gravity, which includes matter and reproduces the Einstein's equations. The dynamical variable in the action is the connection, whose  variation  leads to  the Einstein's equations. This is in contrast with the other approaches in the literature to include matter in Eddington gravity, where the gravitational field equations are different from the Einstein's equations. The variational principle proposed in this work is also well-posed, unlike in the case of the Hilbert action. This is because our action  differs from the Hilbert action in two crucial respects: (a) First,  the gravitational part of the action involves only the connection and has no reference to the metric. (b) Second,  fixing the connection  at the boundary turns out to be  sufficient to render the variational principle well-posed. 

Another remarkable feature of our analysis (which is common to the standard Eddington gravity to a certain extent) is the emergence of a cosmological constant naturally;  more importantly, the cosmological constant \textit{has to be} non-zero for the variational problem to make sense. This fact renders our action to be a \textit{better} choice to derive the Einstein's equations than the Hilbert action, since the latter does not demand a non-zero cosmological constant.

We will now mention several possible extensions of this action principle and their consequences:

To begin with, it is possible to construct a more general class of actions and still obtain the standard Einstein's equations, along the lines of how we have proceeded. One such class of actions can be constructed as follows: Let 
$M_{ab}=L^2[R_{ab}-\kappa T_{ab}]$, where $L$ is a constant length scale introduced for dimensional reasons (which does not affect the variation or the equations of motion) and let $X\equiv |\textrm{det.}M|/|\textrm{det.}g|$ be the ratio of the two determinants, which will transform as a scalar under coordinate transformations. We take the Lagrangian to be an \textit{arbitrary} scalar function of $X$ so that the (dimensionless) action for the gravity + matter system is given by:
\begin{align}\label{fofxaction1}
\mathcal{A}=\int \frac{d^{d}x}{L^d}~\sqrt{|\textrm{det.}g|}~f\left(\frac{|\textrm{det.}M|}{|\textrm{det.}g|}\right)
=\int \frac{d^{d}x}{L^d}~\sqrt{|\textrm{det.}g|}~f(X)~.
\end{align}
The choice of $f(X)=X$ reduces this action to the one in \ref{action}. For other choices of $f(X)$, even the gravitational sector has a dependence in the metric; however, we treat the connection and the metric as independent and vary \textit{only} the connection in the action. It is shown in \ref{AppE} that the variation of the action in \ref{fofxaction1} also leads to Einstein's equations.

Second, let us consider the possible origin of the action in \ref{action}. The occurrence of a determinant in the Lagrangian is strongly suggestive of a path integral origin. To make this connection precise, consider the standard result of a Gaussian path integral in $d=4$ Euclidean space, leading to an effective action:
\begin{align}\label{tppi}
 \int \mathcal{D}v^{a}~\exp\left[-\int \frac{d^4 x}{L^4}\sqrt{|\textrm{det.}(g)|}~v^a (L^2M_{ab}) v^b \right]
 &\propto \exp\left[-\frac{1}{2}\int \frac{d^4 x}{L^4}\sqrt{|\textrm{det.}(g)|}~\ln \left(|\textrm{det.}(L^2 M_{ab})|\right)\right]\nonumber\\
 &\propto \exp\left(-\mathcal{A}_{\rm eff}[M_{ab}]\right)~,
\end{align} 
where, $v^{a}$ is a vector field that is integrated out and $M_{ab}=R_{(ab)}(\Gamma)-\kappa \bar T_{ab}$, is as defined earlier. Here $R_{(ab)}$ is the symmetric part of the Ricci tensor and is treated as a functional of the connection. The $L$ is a constant length scale introduced purely for dimensional reasons (which will be of the order of Planck length in the emergent paradigm). The path integral thus gives rise to the following effective action: 
\begin{equation}\label{logdetact}
\mathcal{A}_{\rm eff}=\frac{1}{2}\int \frac{d^4 x}{L^4}\sqrt{|\textrm{det.}(g)|}~\ln \left(|\textrm{det.}[L^2 (R_{ab}(\Gamma)-\kappa \bar T_{ab})]|\right)~,
\end{equation} 
which is very similar to our action in \ref{action} except for the logarithmic dependence. The variation of this action \textit{also} leads to the Einstein's equations, as shown in \ref{AppE2}. (The action in \ref{logdetact} is equivalent to the one in \ref{fofxaction1} for the choice $f(X)=\ln X$, when we use the fact that the metric dependent terms are not varied in the action. So the result follows from that for the class of actions in \ref{fofxaction1}.)
The interpretation of this analysis and its connection with the microscopic degrees of freedom on null surfaces will be explored in a separate publication \cite{TPSC2020}. 

We conclude by pointing out a key, broader,  implication of the results in this paper, including those of this section, for quantum gravity. As far as classical theories are concerned, it is only the equations of motion that are relevant. The action principle is more of an exercise in elegance and economy and, of course, is the simplest route to incorporate the expected symmetries of the theory. The situation, however, is quite different in quantum theory. In the path integral formalism, for example, it is important to know the form of the action principle as well as the status of dynamical variables. We have now shown that one can obtain Einstein's equations from different choices of action functionals and dynamical variables. (Recall that we only varied the connection and kept the metric frozen.) All of them are equivalent at the classical level but their quantum versions will be quite different. It is conceivable that some of them will lead to a tractable model for quantum gravity, at least in the matter-free case.

\section*{Acknowledgements}

We thank Krishnamohan Parattu for several rounds of extensive discussions on different aspects of this work and for constructing the derivation given in Appendix B. Research of S.C. is funded by the INSPIRE Faculty fellowship from DST, Government of India (Reg. No. DST/INSPIRE/04/2018/000893) and by the Start-Up Research Grant from SERB, DST, Government of India (Reg. No. SRG/2020/000409). The research of T.P is partially supported by the J.C.Bose Fellowship of the Department of Science and Technology, Government of India.
 
\appendix
\labelformat{section}{Appendix #1} 
\labelformat{subsection}{Appendix #1}
\section{A pedagogical note: Matrices versus tensors}\label{AppA}

In differential geometry we often work with the determinant of second rank tensors. The purpose of this Appendix is to stress certain aspects of this relationship between second rank tensors and matrices, which are not adequately emphasized in the textbooks. We will also point out the special status of metric tensor in this context. We will work with \textit{symmetric} second rank tensors for simplicity.

We begin by noting that the most \textit{natural} placement of indices, while describing an element $M^a_b$ of a matrix $M$, is in the mixed (1,1) form. This is obvious from the fact that we want the elements of the identity matrix $I$ to be mapped to the Kronecker delta, which is also a genuine (1,1) tensor $\delta^a_b$ with the same components in all reference frames. (The (0,2) and (2,0) forms of the Kronecker tensor will be $g^{ab}$ and $g_{ab}$ which, of course, will have different components in different coordinate systems.) Representing a matrix in mixed tensor form also allows the natural operation of summation convention (with summation over repeated indices \textit{with one index appearing as subscript and another as superscript}) through the tensorial relation $M^a_bP^b_c=K^a_c$, translating to the matrix multiplication: $MP=K$. (As a special case, this allows the definition of inverses of matrices by $MN=I$ becoming $M^a_bN^b_c=\delta^a_c$ in tensor language.) The trace of the matrix is defined as the sum of its diagonal elements and this is represented by the tensorial equation  $M^a_a=M^a_b\delta^b_a$; this will be numerically same as the matrix trace $Tr M$. 
The determinant of the matrix, in $d=4$, can be expressed in terms of its individual (1,1) form elements through the relation:
\begin{align}
\left(\textrm{det.}M \right)=(1/4!)\delta^{abcd}\delta_{pqrs}M^{p}_{~a}M^{q}_{~b}M^{r}_{~c}M^{s}_{~d}~.
\label{tp1}
\end{align}
Here $\delta^{abcd}$ and $\delta_{pqrs}$ are both alternating symbols taking value $+1$ for even permutation of $(0,1,2,3)$ and taking value $-1$ for odd permutation of the same. Even though neither of the alternating symbols are tensors --- they are tensor densities --- the product appearing in \ref{tp1} is a tensor. Therefore this equation defines the $(\textrm{det.}M)$ as a generally covariant scalar. The representation of the matrix element in the (1,1) tensor representation is crucial for this. All these show that it is trivial to convert between a (1,1) mixed tensor and a matrix; you do not have to do anything and the components directly map to each other.

But what do we do if we want to express a (0,2) or (2,0) tensor as a matrix and, say, take its determinant or trace? The most important example is the metric tensor $g_{ab}$ written in (0,2) form; how do we find its determinant or trace? 

Let us start with the determinant. If you try to convert the (0,2) form to (1,1) form and \textit{then} define the determinant, you will not get what is usually denoted by $(\textrm{det.}g_{ab})$ because the (1,1) \textit{tensor} corresponding to $g_{ab}$ is $\delta^a_b$! What we have to do --- and is standard practice, though this procedure is not spelt out in such gory detail -- is to establish a correspondence between the, say, (0,2) tensor and a matrix (with the elements of the matrix \textit{always} given in (1,1) form). For example, given some (0,2) tensor $\mathcal{T}_{ab}$ we  \textit{define} a matrix $M^{a}_{~b}$, such that the $(a,b)$th component of the tensor $\mathcal{T}_{ab}$ is mapped to the element at the intersection of the $a$th row and the $b$th column of the matrix $M^{a}_{~b}$. Then we  define the determinant of a second rank tensor $\mathcal{T}_{ab}$, as the determinant of the matrix $M$ with elements  $M^{a}_{~b}$.

For example, given the metric of the Friedman universe as a (0,2) tensor $g_{ab}$ through $ds^2=-dt^2+a^2(t)d\bm{x}^2$. The process described above maps this (0,2) tensor to a diagonal matrix $M^{a}_{b}=\textrm{diag.}(-1,a^2,a^2,a^2)$ having determinant $\textrm{det.}M=-a^{6}$. This is the quantity we \textit{define} to be the determinant of the original metric tensor $g_{ab}$ and take its value to be $\textrm{det.}g=-a^6$. We do not write $g_{ab}$ in the (1,1) form --- getting $\delta^a_b$ --- map it to a matrix and take its determinant (which would have given just $-1$). 

Once we have defined the $\sqrt{-g}$ factor by the above procedure, we can define genuine tensors $\epsilon_{pqrs}=\sqrt{-g}\delta_{pqrs}$ and $\epsilon^{abcd}=-(1/\sqrt{-g})\delta^{abcd}$ in terms of the alternating symbols and use them to define determinants. 
The determinant of a second rank (0,2) tensor $\mathcal{T}_{ab}$ can be obtained by first mapping it to the matrix $M^{a}_{~b}$ introduced above, (with the index placements (1,1)) and then using the result, 
\begin{align}
\left(\textrm{det.}~\mathcal{T}\right)&=(1/4!)\delta^{abcd}\delta_{pqrs}M^{p}_{~a}M^{q}_{~b}M^{r}_{~c}M^{s}_{~d}\to (1/4!)\delta^{abcd}\delta^{pqrs}\mathcal{T}_{ap}\mathcal{T}_{bq}\mathcal{T}_{cr}\mathcal{T}_{ds}~
\nonumber
\\
&=(1/4!)\left(\sqrt{-g}\right)^{2}\epsilon^{abcd}\epsilon^{pqrs}\mathcal{T}_{ap}\mathcal{T}_{bq}\mathcal{T}_{cr}\mathcal{T}_{ds}~.
\end{align}
The first equality comes from the matrix theory, the second from our mapping between matrix elements and the tensor (by definition). 
Note that, the terms involving $\epsilon^{abcd}$ and $\mathcal{T}_{ap}$ form a scalar and hence $\left(\textrm{det.}~\mathcal{T}_{ab}\right)$ is a scalar \textit{density} of weight 1, due to the presence of the $\left(\sqrt{-g}\right)^{2}$ factor. Similarly, for a tensor $\mathcal{S}^{ab}$, of the (2,0) form, the determinant is defined as,
\begin{align}
\textrm{det.} \mathcal{S}=(1/4!)\left(\sqrt{-g}\right)^{-2}\epsilon_{abcd}\epsilon_{pqrs}\mathcal{S}^{ap}\mathcal{S}^{bq}\mathcal{S}^{cr}\mathcal{S}^{ds}~,
\end{align}
and hence the determinant of a tensor $\mathcal{S}^{ab}$ is a scalar density of weight -1, as desired. (In the case of (1,1) tensor, in \ref{tp1}, the two $\sqrt{-g}$ factors cancel and we get a scalar, with just the alternating symbol.)

Once a tensor has been mapped to a matrix $M^{a}_{~b}$, its inverse matrix $N^{a}_{~b}$ is defined as the matrix, which satisfies the following relations, $M^{a}_{~b}N^{b}_{~c}=\delta^{a}_{c}$ and $N^{a}_{~b}M^{b}_{~c}=\delta^{a}_{c}$. This allows us to define another \textit{tensor} $\mathcal{S}^{ab}$, by the correspondence, $N^{a}_{~b}\leftrightarrow \mathcal{S}^{ab}$. That is, the element at the intersection of the $a$th row and $b$th column of $N^{a}_{~b}$ is mapped to the $(a,b)$th element of $\mathcal{S}^{ab}$. This leads to the following relation between $\mathcal{T}_{ab}$ and $\mathcal{S}^{ab}$ treated as tensors: $\mathcal{T}_{ab}\mathcal{S}^{bc}=\delta^{c}_{a}=\mathcal{S}^{cb}\mathcal{T}_{ba}$. Note that existence of the tensor $\mathcal{S}^{ab}$ does not require a metric and can be arrived by just inverting the matrix associated with $\mathcal{T}_{ab}$. We have used  these results in the main text. 

We conclude with a few comments on the Trace operation. One possibility is to define the trace of any second rank tensor $\mathcal{T}_{ab}$, by mapping it to the matrix $M^{a}_{~b}$ and then computing its trace. This will be consistent with the expression for the determinant of the second rank tensor $\mathcal{T}_{ab}$, such that,
\begin{align}\label{tp4}
\textrm{det}. \left(\mathcal{T}\right)=\textrm{det.} \left(M\right)=\exp\left(\textrm{Tr} \ln M^{a}_{~b}\right)~.
\end{align}
(This relation can be derived by going to the diagonal basis and noting that $\textrm{Tr}\ln M^{a}_{~b}=\sum_{r}\ln \lambda_{r}$, where $\lambda_{r}$ are the eigenvalues of the matrix $M^{a}_{~b}$.) Note that the trace operation will act on the logarithm of the matrix $M^{a}_{b}$ and is \emph{not} equal to the tensor trace operation, $g^{ab}\mathcal{T}_{ab}$. A familiar example is the metric tensor in 3-dimensional spherical polar coordinates, which we map to the matrix dia $[1, r^2, r^2\sin^2\theta]$  so that we get the correct determinant $\textrm{det.}g=r^{4}\sin^{2}\theta$. The trace of the logarithm of this matrix $M^{a}_{b}$ --- which we have used to define the \textit{determinant} of the metric ---  will be $\ln (1)+\ln (r^{2})+\ln (r^{2}\sin^{2}\theta)=4\ln (r)+2\ln (\sin\theta)$. This result, when substituted in the above equation, will yield $\textrm{det.}g=r^{4}\sin^{2}\theta$, which is indeed the correct expression. But if one defines the tensorial trace, which for an \textit{arbitrary} tensor $Q_{ab}$ is of the form: $Q=g^{ab}Q_{ab}$, and use it for the metric tensor of spherically symmetric spacetime, we will always get $g^{ab}g_{ab}=3$, which will not, of course,  yield the appropriate expression for the \textit{determinant} of the metric tensor through \ref{tp4}. Similarly, if we consider a three dimensional tensor $\mathcal{T}_{ab}$, which has only three non-zero entries, $\mathcal{T}_{11}=t_{1}$, $\mathcal{T}_{22}=t_{2}$ and $\mathcal{T}_{33}=t_{3}$, the associated matrix will be $M^{a}_{~b}=\textrm{diag.}(t_{1},t_{2},t_{3})$ and the matrix trace is given by $\textrm{Tr}~M^{a}_{~b}=t_{1}+t_{2}+t_{3}$, while the tensor trace will correspond to, $T^{a}_{a}=g^{11}t_{1}+g^{22}t_{2}+g^{33}t_{3} \neq \textrm{Tr}~M^{a}_{~b}$. Thus in general, the tensorial trace of a (0,2) (or, (2,0)) second rank tensor is not equal to the trace of a matrix to which the tensor has been mapped to, in the sense described above. 

\section{Proof of the identity presented in \ref{identity}}\label{AppB}

We will give the proof of the identity  in \ref{identity}, which can be obtained by  several methods. We will present the most relevant one here, wherein we shall start by deriving the covariant derivative of a scalar density of weight $-1$. 

Let $A_{ab}$ be an arbitrary tensor with two covariant indices, having a matrix representation as described in \ref{AppA}. Then, with $\det(A_{ab})$ denoting the determinant of the matrix, originating from the tensor $A_{ab}$, it follows that $\det(A_{ab})=\det(A_{ba})$. Since no further confusion is likely to arise, we will drop the subscripts and write it just as $\det(A)$; it follows that  $\sqrt{\det(A)}$ is a scalar density of weight $-1$. Following the discussions in \ref{AppA}, we also define the tensor corresponding to  the inverse matrix as $B^{ab}$, such that $B^{ac}A_{cb}=\delta^a_b=A_{bc}B^{ca}$, where we have used the fact that the left inverse of a matrix is the same as its right inverse. (As we have already emphasized, $B^{ab}$ is \emph{not} equal to $A^{ab}$, obtained by raising the indices of $A_{ab}$ by the metric tensor.) Keeping this mind, and taking the covariant derivative of $\sqrt{\det(A)}$, we obtain
\begin{align}
\nabla_{a} [\sqrt{\det(A)}]&= \frac{\sqrt{\det(A)}}{2}B^{bc}\nabla_{a} A_{cb}= \frac{\sqrt{\det(A)}}{2}B^{bc}\left(\partial_a A_{cb}-\Gamma^{d}_{ac}A_{db}-\Gamma^{d}_{ab}A_{cd}\right) 
\nn
\\
&=\partial_{a} [\sqrt{\det(A)}]-\frac{\sqrt{\det(A)}}{2}B^{bc}\left(\Gamma^{d}_{ac}A_{db}+\Gamma^{d}_{ab}A_{cd}\right) 
\nn
\\
&=\partial_{a} [\sqrt{\det(A)}]-\frac{\sqrt{\det(A)}}{2}\left(\Gamma^{d}_{ac}A_{db}B^{bc}+\Gamma^{d}_{ab}B^{bc}A_{cd}\right) 
\nn
\\
&=\partial_{a} [\sqrt{\det(A)}]-\frac{\sqrt{\det(A)}}{2}\left(\Gamma^{d}_{ac}\delta_{d}^{c}+\Gamma^{d}_{ab}\delta_{d}^{b}\right) 
\nn
\\
&=\partial_{a} [\sqrt{\det(A)}]-\frac{\sqrt{\det(A)}}{2}\left(\Gamma^{c}_{ac}+\Gamma^{b}_{ab}\right)~.
\end{align}
Thereby implying the following identity,
\begin{align}\label{cov_density1}
\nabla_{a} [\sqrt{\det(A)}]&=\partial_{a} [\sqrt{\det(A)}]-\Gamma^{b}_{ab}\sqrt{\det(A)}~.
\end{align}
This is the basic result we need, which, of course, should be familiar in the context of the tensor $A_{ij}$ being the metric tensor $g_{ij}$. (In that case,  it is customary to take the negative of the determinant inside the square root and work with $\sqrt{-\det(g)}$. This is to ensure that one always obtains a positive quantity inside the square root. Since we are just proving an algebraic formula for a general tensor, we shall not bother to do this.) Finally, using \ref{cov_density1}, we can derive the desired result, 
\begin{align}
\nabla_{c}\left[\sqrt{\textrm{det.}\left(M\right)}~V^{c}\right]&=\sqrt{\textrm{det.}\left(M\right)}\left(\partial_{c}V^{c}+\Gamma^{c}_{cd}V^{d}\right)+V^{c}\nabla_{c}\sqrt{\textrm{det.}\left(M\right)} 
\nn
\\
&=\sqrt{\textrm{det.}\left(M\right)}\left(\partial_{c}V^{c}+\Gamma^{c}_{cd}V^{d}\right)+V^{c}\left(\partial_{c}\sqrt{\textrm{det.}\left(M\right)}-\Gamma^{d}_{cd}\sqrt{\det.(M)}\right)
\nn
\\
&=\partial_{c}\left[\sqrt{\textrm{det.}\left(M\right)}~V^{c}\right]~,
\end{align}
which we have used in the main text. 

\section{Eddington gravity with matter and torsion}\label{AppC}

In this section, we will describe the inclusion of torsion \cite{Hehl:1976kj,Hehl:1974cn,Chakraborty:2018qew} in the framework presented in this work, where the matter degrees of freedom have been incorporated within the Eddington gravity paradigm. Given the action presented in \ref{action} in the torsionless case, one may consider the following Lagrangian, in the presence of torsion, $L=[|\textrm{det.}(\bar{R}_{(ab)}-\bar{T}_{ab})|]^{1/2}$, where, $\bar{R}_{ab}$ depends on the spacetime torsion and has the following expression,
\begin{align}
\bar{R}_{ab}=R_{ab}+\bar{\nabla}_{c}K^{c}_{~ab}-\bar{\nabla}_{a}K^{c}_{~cb}-K^{c}_{~cd}K^{d}_{~ab}+K^{c}_{~da}K^{d}_{~cb}~.
\end{align}
Here, $K^{c}_{~ab}$ is the contorsion tensor, with the symmetry property being: $K_{cab}=-K_{bac}$. The torsion tensor, on the other hand, is defined as, $T^{c}_{~ab}\equiv K^{c}_{~ab}-K^{c}_{~ba}$ with the symmetry property, $T^{c}_{~ab}=-T^{c}_{~ba}$. Further, we have, $K^{c}_{~cd}=T^{c}_{~cd}=-T_{d}$, denoting the trace of the torsion tensor. Thus we obtain,
\begin{align}
\bar{R}_{ab}&=R_{ab}+\bar{\nabla}_{c}K^{c}_{~ab}+\bar{\nabla}_{a}T_{b}+T_{d}K^{d}_{~ab}+K^{c}_{~da}K^{d}_{~cb}~,
\end{align}
Note that this tensor $\bar{R}_{ab}$ is not symmetric and hence in the action we had to include the symmetric part of the same. As evident,  the variation of the action due to the variation of the connection, while keeping the contorsion tensor fixed, is solely due to the variation of the Ricci tensor $\bar{R}_{ab}$. After neglecting appropriate boundary terms, it turns out that the variation of the action, does not lead to the appropriate Einstein-Cartan field equations. Thus the simplest extension of the action presented in \ref{action} will not work in the presence of torsion. 

Rather, we should consider the following form for the action, in the presence of torsion,
\begin{align}\label{action_new}
\mathcal{A}_{\rm torsion}=\int d^{4}x\sqrt{|\textrm{det.}\left[\bar{R}_{(ab)}-\left(\bar{\nabla}_{c}+T_{c}\right)\left(K^{c}_{~(ab)}+\delta^{c}_{(a}T_{b)}\right)-\bar{T}_{ab}\right]|}~,
\end{align}
where, $\bar{R}_{ab}$ and $\bar{T}_{ab}$ have their usual meaning, as defined earlier. Defining, $M_{ab}$ as the symmetric tensor within the determinant presented in \ref{action_new} and using the expression for $\bar{R}_{ab}$, we obtain,
\begin{align}
M_{ab}&\equiv\bar{R}_{(ab)}-\left(\bar{\nabla}_{c}+T_{c}\right)\left(K^{c}_{~(ab)}+\delta^{c}_{(a}T_{b)}\right)-\bar{T}_{ab}
\nonumber
\\
&=R_{(ab)}+\bar{\nabla}_{c}K^{c}_{~(ab)}+\bar{\nabla}_{(a}T_{b)}+T_{d}K^{d}_{~(ab)}+K^{c}_{~d(a}K^{d}_{~cb)}-\left(\bar{\nabla}_{c}+T_{c}\right)\left(K^{c}_{~(ab)}+\delta^{c}_{(a}T_{b)}\right)-\bar{T}_{ab}
\nonumber
\\
&=R_{(ab)}-T_{(a}T_{b)}+K^{c}_{~d(a}K^{d}_{~cb)}-\bar{T}_{ab}~.
\end{align}
Thus variation of the action due to the variation of the connection, keeping the contorsion tensor fixed, is related to the variation of $M_{ab}$ due to $\Gamma^{c}_{ab}$, yielding, 
\begin{align}
\delta_{\Gamma} \mathcal{A}_{\rm new}&=-\frac{1}{2}\int d^{d}x~\nabla_{d}\left[\sqrt{|\textrm{det.}\left(M\right)|}~\delta^{cd}_{ab}N^{pa}\right]\delta \Gamma^{b}_{cp}~,
\end{align}
where a total derivative term has been neglected. As we have noticed in the main text, the solution to the above equation of motion being, $M_{ab}=\lambda g_{ab}$, yielding the appropriate Einstein-Cartan gravitational field equations. While variation of the action with respect to the contorsion tensor, keeping the connection fixed, yields,
\begin{align}
\delta_{K} \mathcal{A}_{\rm new}&=\frac{1}{2}\int d^{d}x~\sqrt{|\textrm{det.}\left(M\right)|}~\delta K^{d}_{~cb}\Bigg\{2N^{ba}K^{c}_{~da}+2N^{ba}\delta^{c}_{d}T_{a}\Bigg\}~.
\end{align}
Thus setting the variation of the action to zero for arbitrary variation of the contorsion tensor and with $N^{ab}=\lambda^{-1}g^{ab}$, we observe that, the above equation of motion demands, $K^{a}_{~cd}+\delta^{a}_{c}T_{d}=0$, which yields, $K^{a}_{~cd}=0$. Hence the contorsion tensor identically vanishes, as it should in absence of any source for the contorsion tensor in the matter sector.   

\section{Eddington-type action for pure \LL\ gravity}\label{AppD}

In this appendix, taking a cue from the derivation of the Einstein's equations, we will derive appropriate field equations for pure Lovelock gravity \'{a} la Eddington. For this purpose, we start with the following action:
\begin{align} 
\mathcal{A}&=\int d^{d}x~\sqrt{|\textrm{det.}\left(\mathcal{R}^{(m)}_{(ab)}-\kappa \bar{T}^{(m)}_{ab}\right)|}~;\qquad \bar{T}^{(m)}_{ab}=T_{ab}-\frac{1}{(d-2m)}Tg_{ab}~;
\nonumber
\\
\mathcal{R}^{(m)}_{ab}&=P_{a}^{~pqr}R_{bpqr}=m~\delta^{c_{1}d_{1}c_{2}d_{2}\cdots c_{m}d_{m}}_{a_{1}b_{1}a_{2}b_{2}\cdots a_{m}b_{m}}\left(g^{b_{2}e_{2}}R^{a_{2}}_{~~e_{2}c_{2}d_{2}}\cdots g^{b_{m}e_{m}}R^{a_{m}}_{~~e_{m}c_{m}d_{m}} \right)\delta^{a_{1}}_{a}g^{b_{1}e_{1}}g_{bq}R^{q}_{~e_{1}c_{1}d_{1}}~.
\end{align}
The variation of the above action with respect to the connection (keeping the metric fixed) gives:
\begin{align}
\delta _{\Gamma}\mathcal{R}^{(m)}_{ab}&=m^{2}~\delta^{c_{1}d_{1}c_{2}d_{2}\cdots c_{m}d_{m}}_{a_{1}b_{1}a_{2}b_{2}\cdots a_{m}b_{m}}\left(g^{b_{2}e_{2}}R^{a_{2}}_{~~e_{2}c_{2}d_{2}}\cdots g^{b_{m}e_{m}}R^{a_{m}}_{~~e_{m}c_{m}d_{m}} \right)\delta^{a_{1}}_{a}g^{b_{1}e_{1}}g_{bq}\left(\delta_{\Gamma}R^{q}_{~e_{1}c_{1}d_{1}}\right)
\nonumber
\\
&=m^{2}~\delta^{c_{1}d_{1}c_{2}d_{2}\cdots c_{m}d_{m}}_{a_{1}b_{1}a_{2}b_{2}\cdots a_{m}b_{m}}\left(g^{b_{2}e_{2}}R^{a_{2}}_{~~e_{2}c_{2}d_{2}}\cdots g^{b_{m}e_{m}}R^{a_{m}}_{~~e_{m}c_{m}d_{m}} \right)\delta^{a_{1}}_{a}g^{b_{1}e_{1}}g_{bq}\left(\nabla_{c_{1}}\delta \Gamma^{q}_{e_{1}d_{1}}-\nabla_{d_{1}}\delta \Gamma^{q}_{e_{1}c_{1}}\right)
\nonumber
\\
&=2m^{2}~\delta^{c_{1}d_{1}c_{2}d_{2}\cdots c_{m}d_{m}}_{a_{1}b_{1}a_{2}b_{2}\cdots a_{m}b_{m}}\left(g^{b_{2}e_{2}}R^{a_{2}}_{~~e_{2}c_{2}d_{2}}\cdots g^{b_{m}e_{m}}R^{a_{m}}_{~~e_{m}c_{m}d_{m}} \right)\delta^{a_{1}}_{a}g^{b_{1}e_{1}}g_{bq}\left(\nabla_{c_{1}}\delta \Gamma^{q}_{e_{1}d_{1}}\right)
\nonumber
\\
&=2m~P^{c_{1}d_{1}}_{a_{1}b_{1}}\delta^{a_{1}}_{a}g^{b_{1}e_{1}}g_{bq}\left(\nabla_{c_{1}}\delta \Gamma^{q}_{e_{1}d_{1}}\right)
\end{align}
Therefore, defining, $M_{ab}\equiv \mathcal{R}^{(m)}_{(ab)}-\kappa \bar{T}^{(m)}_{ab}$ and its inverse as $N^{ab}$, both of which are symmetric, the above variation of the action becomes,
\begin{align} 
\delta_{\Gamma}\mathcal{A}&=\int d^{d}x~\sqrt{|\textrm{det.}\left(M\right)|}~N^{ab}\delta_{\Gamma}\mathcal{R}^{(m)}_{(ab)}
\nonumber
\\
&=2m~\int d^{d}x~\sqrt{|\textrm{det.}\left(M\right)|}~P^{c_{1}d_{1}}_{a_{1}b_{1}}\delta^{a_{1}}_{a}g^{b_{1}e_{1}}g_{bq}N^{ab}\left(\nabla_{c_{1}}\delta \Gamma^{q}_{e_{1}d_{1}}\right)
\nonumber
\\
&=2m~\int d^{d}x~\nabla_{c_{1}}\left(\sqrt{|\textrm{det.}\left(M\right)|}~P^{~e_{1}c_{1}d_{1}}_{a_{1}}\delta^{a_{1}}_{a}g_{bq}N^{ab}\delta \Gamma^{q}_{e_{1}d_{1}}\right)
\nonumber
\\
&\hskip 2 cm - 2m~\int d^{d}x~\nabla_{c_{1}}\left(\sqrt{|\textrm{det.}\left(M\right)|}~P^{c_{1}d_{1}}_{a_{1}b_{1}}\delta^{a_{1}}_{a}g^{b_{1}e_{1}}g_{bq}N^{ab}\right)\delta \Gamma^{q}_{e_{1}d_{1}}
\end{align}
The quantity, $P^{~e_{1}c_{1}d_{1}}_{a_{1}}\delta^{a_{1}}_{a}g_{bq}N^{ab}\delta \Gamma^{q}_{e_{1}d_{1}}$ transforms as a vector and hence by the identity presented in \ref{AppB}, it follows that the first term is a total divergence. Discarding this boundary contribution, we obtain the following field equation from the second term, by setting the variation of the action to be zero, for arbitrary variation of the connection,
\begin{align} 
\left[\nabla_{c_{1}}\left(\sqrt{|\textrm{det.}\left(M\right)|}~P^{c_{1}d_{1}}_{a_{1}b_{1}}\delta^{a_{1}}_{a}g^{b_{1}e_{1}}g_{bq}N^{ab}\right)+\nabla_{c_{1}}\left(\sqrt{|\textrm{det.}\left(M\right)|}~P^{c_{1}e_{1}}_{a_{1}b_{1}}\delta^{a_{1}}_{a}g^{b_{1}d_{1}}g_{bq}N^{ab}\right)\right]=0
\end{align}
Since, $\nabla_{c_{1}}P^{c_{1}d_{1}}_{a_{1}b_{1}}=0$ for pure Lovelock theories \cite{Dadhich:2008df,Dadhich:2012zq,Padmanabhan:2013xyr,Chakraborty:2015wma}, it follows that the above equation can be expressed as,
\begin{align} 
\left[P^{c_{1}d_{1}}_{a_{1}b_{1}}\nabla_{c_{1}}\left(\sqrt{|\textrm{det.}\left(M\right)|}~\delta^{a_{1}}_{a}g^{b_{1}e_{1}}g_{bq}N^{ab}\right)+P^{c_{1}e_{1}}_{a_{1}b_{1}}\nabla_{c_{1}}\left(\sqrt{|\textrm{det.}\left(M\right)|}~\delta^{a_{1}}_{a}g^{b_{1}d_{1}}g_{bq}N^{ab}\right)\right]=0
\end{align}
This equation has the first integral, given by $M_{ab}=\lambda g_{ab}$, leading to the compatibility between the connection and the metric as well as the field equations:
\begin{align} 
\mathcal{R}_{(ab)}^{(m)}-\kappa \bar{T}^{(m)}_{ab}=\lambda g_{ab}~.
\end{align}
Taking trace of this equation, we obtain, 
\begin{align}
\frac{\kappa}{(d-2m)}T=\frac{d}{2m}\lambda-\frac{1}{2}L_{(m)}~,
\end{align}
where, $L_{(m)}\equiv (1/m)P^{abcd}R_{abcd}$ is the Lovelock Lagrangian in the standard metric formalism. Thus the above higher curvature action, involving the determinant of the Ricci-equivalent term, yields the standard gravitational field equations for pure Lovelock gravity:
\begin{equation}
 \mathcal{R}_{(ab)}^{(m)}-\frac{1}{2}L_{(m)}g_{ab}+\Lambda g_{ab}=\kappa T_{ab}; \qquad \Lambda=\left(\frac{d-2m}{2m}\right)\lambda
\end{equation}
Note that for $d=2m$, i.e., in critical dimensions for Lovelock gravity, the effect from the cosmological constant term identically vanishes.

\section{A more general form of the action}\label{AppE}

It turns out that the action used in this paper, given in \ref{action} is just the simplest one amongst a very general class of actions, which will lead to Einstein's field equations along the lines discussed in the text. Here we will illustrate this for one such class of actions. 

Let us again define, as in the main text, $M_{ab}=L^2[R_{ab}-\kappa T_{ab}]$, and let $X\equiv |\textrm{det.}M|/|\textrm{det.}g|$ be the ratio of the two determinants, which will transform as a scalar under coordinate transformations. (The constant length scale $L$ is introduced just for dimensional reasons and does not affect the derivation.) We take the Lagrangian to be an \textit{arbitrary} scalar function of $X$ so that the action for the gravity + matter system is given by:
\begin{align}\label{fofxaction}
\mathcal{A}=\int \frac{d^{d}x}{L^d}~\sqrt{|\textrm{det.}g|}~f\left(\frac{|\textrm{det.}M|}{|\textrm{det.}g|}\right)
=\int \frac{d^{d}x}{L^d}~\sqrt{|\textrm{det.}g|}~f(X).
\end{align}
which is a scalar. The action in \ref{action}  arises for the simplest choice of $f$ viz., $f(X)=X$, Just as in the main text, we will treat $R_{ab}(\Gamma)$ to be a function of the connection, treat the connection and the metric to be independent and vary only the connection in the action.
Thus, for arbitrary variation of the connection, the variation of the above action becomes, exactly as in the main text (with the ``inverse'' $N^{ab}$ defined by the relation $N^{ab}M_{bc}=\delta^a_c$):
\begin{align}
\delta_{\Gamma}\mathcal{A}&=\int \frac{d^{d}x}{L^d}~\sqrt{|\textrm{det.}g|}~f'(X)\frac{\delta |\textrm{det.}M|}{|\textrm{det.}g|}
\nonumber
\\
&=\int \frac{d^{d}x}{L^d}~\sqrt{|\textrm{det.}g|}~Xf'(X)N^{ab}\delta M_{ab}
\nonumber
\\
&=\int \frac{d^{d}x}{L^d}~\sqrt{|\textrm{det.}g|}~Xf'(X)\delta^{cd}_{ab}N^{pa}\nabla_{d}\delta \Gamma^{b}_{cp} 
\nonumber
\\
&=\int \frac{d^{d}x}{L^d}~\nabla_{d}\left[\sqrt{|\textrm{det.}g|}~Xf'(X)\delta^{cd}_{ab}N^{pa}\delta \Gamma^{b}_{cp} \right]
\nonumber
\\
&\hskip 1 cm -\int \frac{d^{d}x}{L^d}~\nabla_{d}\left[\sqrt{|\textrm{det.}g|}~Xf'(X)\delta^{cd}_{ab}N^{pa}\right]\delta \Gamma^{b}_{cp} 
\end{align}
The quantity, $Xf'(X)\delta^{cd}_{ab}N^{pa}\delta \Gamma^{b}_{cp}$ transforms as a vector under coordinate transformation and hence the first term in the variation of the above action is a total derivative term, which can be converted to the boundary term. This term does not contribute since we will  set $\delta\Gamma^b_{cp}=0$ at the boundary. Thus the field equation becomes, on taking into account the symmetry of $\delta\Gamma^b_{cp}$ on $c,p$:
\begin{align}
\nabla_{d}\left[\sqrt{|\textrm{det.}g|}~Xf'(X)\delta^{cd}_{ab}N^{pa}\right]+\nabla_{d}\left[\sqrt{|\textrm{det.}g|}~Xf'(X)\delta^{pd}_{ab}N^{ca}\right]=0
\end{align}
which, when expanded, yields,
\begin{align}
2\nabla_{b}\left[\sqrt{|\textrm{det.}g|}~Xf'(X)N^{cp}\right]&-\nabla_{a}\left[\sqrt{|\textrm{det.}g|}~Xf'(X)N^{pa}\right]\delta^{c}_{b}
\nonumber
\\
&-\nabla_{a}\left[\sqrt{|\textrm{det.}g|}~Xf'(X)N^{ca}\right]\delta^{p}_{b}=0~.
\end{align}
Multiplying the above expression by $\delta^{b}_{c}$, we get $\nabla_{a}\left[\sqrt{|\textrm{det.}g|}~Xf'(X)N^{ca}\right]=0$ which, when substituted back leads to the result:
\begin{align}\label{fegen}
\nabla_{b}\left[\sqrt{|\textrm{det.}g|}~Xf'(X)N^{cp}\right]=0~.
\end{align}
This requires, $N^{ab}\propto g^{ab}$ thereby making $M_{ab}\propto g_{ab}$ and $X=$ constant. This leads to the Einstein's equations. We illustrate these details in the two examples discussed below: 

\subsection{The case of $f(X)=X^\alpha$}\label{AppE1}

As a first example of the general case discussed above, in this section we choose $f(X)=X^{\alpha}$, a power law with real $\alpha$. Thus we obtain the following action principle (choosing $d=4$ for simplicity):
\begin{align}
\mathcal{A}=\int d^{4}x~\left(|\textrm{det.} g|\right)^{-\alpha+\frac{1}{2}}\left(|\textrm{det.} M|\right)^{\alpha}~.
\end{align}
(We have not explicitly displayed here a length scale $L$ which could be used to make all the variables dimensionless.) Following the variation of the general Lagrangian presented above, from \ref{fegen}, we obtain the following field equation:
\begin{align}
\nabla_{b}\left[\left(|\textrm{det.} g|\right)^{-\alpha+\frac{1}{2}}\left(|\textrm{det.} M|\right)^{\alpha}~N^{pc}\right]=0
\end{align}
which is solved by the following choice:
\begin{align}\label{tp2}
\left(|\textrm{det.} g|\right)^{-\alpha+\frac{1}{2}}\left(|\textrm{det.} M|\right)^{\alpha}~N^{pc}=\lambda \sqrt{|\textrm{det.} g|}g^{pc}
\end{align}
Taking the determinant of both sides, we get $\left(|\textrm{det.} g|\right)^{-4\alpha+2}\left(|\textrm{det.} M|\right)^{4\alpha-1}=\lambda^{4}|\textrm{det.} g|$, which yields, $\left(|\textrm{det.} M|\right)^{4\alpha-1}=\lambda^{4}\left(|\textrm{det.} g|\right)^{4\alpha-1}$. Thus we must have, $M_{ab}=\lambda^{1/(4\alpha-1)}g_{ab}$. This, result, when substituted back into \ref{tp2}, leads to $N^{pc}\propto g^{pc}$, which leads to the Einstein's equations.

\subsection{The case of $f(X)=\ln X$: Effective action arising from a path integral}\label{AppE2}

This case is special for two reasons. First of all, notice that the explicit $X$ dependence in the field equation in \ref{fegen}, contained in the factor $Xf'(X)$ goes away when $f(X)=\ln X$. Second, as mentioned in the main text (see \ref{tppi}), the log-det structure arises very naturally in effective actions defined through path integrals. For this choice of $f(X)$, the action reduces to:
\begin{align}
\mathcal{A}=\int d^{4}x\sqrt{|\textrm{det.} g|}~\ln \left(\frac{|\textrm{det.} M|}{|\textrm{det.} g|}\right)\to
\int d^{4}x\sqrt{|\textrm{det.} g|}~\ln \left(|\textrm{det.} M|\right)~.
\end{align}
The first form of the action explicitly shows the scalar nature of the Lagrangian as the ratio of two determinants. The second form of the action is equivalent to the first since we are not varying the metric. (Again, we have not explicitly displayed here a length scale $L$ which could be used to make all the variables dimensionless.)

The variation of the first form of that action leads, via \ref{fegen}, to $\nabla_b(\sqrt{|\textrm{det.} g|}~N^{pc})=0$ leading, again, to $ N^{pc}\propto g^{pc}$ and thus to the Einstein's equations. Considering the importance of the second form of the action --- which is what a path integral directly leads to --- we will provide an explicit demonstration of this result.

For this purpose, we start with the following form for the action,
\begin{align}
\mathcal{A}=
\int d^{4}x\sqrt{|\textrm{det.} g|}~\ln \left(|\textrm{det.} M|\right)~,
\end{align}
and, as usual, vary the connection. This gives:
\begin{align}
\delta \mathcal{A}&=\int d^{4}x\sqrt{|\textrm{det.} g|}\frac{1}{\left(|\textrm{det.} M|\right)}\delta \left(|\textrm{det.} M|\right)
\nonumber
\\
&=\int d^{4}x\sqrt{|\textrm{det.} g|}~N^{ba}\delta M_{ab}=\int d^{4}x\sqrt{|\textrm{det.} g|}~\delta_{ab}^{cd}~N^{pa}\nabla_{d}\delta \Gamma^{b}_{cp}
\nonumber
\\
&=\int d^{4}x~\nabla_{d}\left\{\sqrt{|\textrm{det.} g|}~\delta_{ab}^{cd}~N^{pa} \delta \Gamma^{b}_{cp}\right\}-\int d^{4}x\nabla_{d}\left\{\sqrt{|\textrm{det.} g|}~\delta_{ab}^{cd}~N^{pa}\right\}\delta \Gamma^{b}_{cp}~.
\end{align}
The first term is a total divergence, due to the identity presented in \ref{AppB}, and does not contribute when the boundary conditions are used. The second term gives the following field equation:
\begin{align}
\nabla_{d}\left\{\sqrt{|\textrm{det.} g|}\delta_{ab}^{cd}~N^{pa}\right\}&+\nabla_{d}\left\{\sqrt{|\textrm{det.} g|}\delta_{ab}^{pd}~N^{ca}\right\}
\nonumber
\\
&=2\nabla_{b}\left(\sqrt{|\textrm{det.} g|}N^{pc}\right)-\delta^{c}_{b}\nabla_{d}\left(\sqrt{|\textrm{det.} g|}N^{pd}\right)-\delta^{p}_{b}\nabla_{d}\left(\sqrt{|\textrm{det.} g|}N^{cd}\right)=0~.
\end{align}
Contraction on the indices $(b,c)$ in the above expression yields, $\nabla_{d}(\sqrt{|\textrm{det.} g|}N^{cd})=0$, which when substituted back in the above expression, provides the following result,
\begin{align}
\nabla_{b}\left(\sqrt{|\textrm{det.} g|}~N^{pc}\right)=0~.
\end{align}
As usual, the above equation is solved by the following choice, $N^{ab}=\lambda^{-1}g^{ab}$, where, $g_{ab}$ is the metric compatible with the connection $\Gamma^{a}_{bc}$. This is equivalent to $M_{ab}=\lambda g_{ab}$, leading to, \ref{Einstein_eq}, which are the Einstein's equations. We will explore the implications of this Lagrangian involving Logarithm of the determinant of $M_{ab}$ in a separate publication \cite{TPSC2020}. 

\bibliography{References}

\bibliographystyle{./utphys1}
\end{document}